\documentclass[a4paper,superscriptaddress,showpacs,prb,twocolumn]{revtex4}
\usepackage{natbib}

\usepackage{amsmath}
\usepackage{graphicx}
\renewcommand{\vec}[1]{\mathbf{#1}}
\newcommand{\chalmersTF}{Department of Applied Physics,
Chalmers University of Technology,
SE-412\;96 G\"{o}teborg, Sweden}
\newcommand{\chalmersMC}{Department of Microtechnology and Nanoscience, MC2,
Chalmers University of Technology,
SE-412\;96 G\"{o}teborg, Sweden}
\newcommand{\camp}{CAMD, Department of Physics,
Technical University of Denmark,
DK-2800 Kgs.~Lyngby, Denmark}

\begin{document}

   \preprint{Applied Physics Report no. 2008-X}

 \author{Jesper Kleis}\affiliation{\chalmersTF}\affiliation{\camp}
  \author{Elsebeth Schr\"oder}\affiliation{\chalmersTF}
 \author{Per Hyldgaard}\email{hyldgaar@chalmers.se}\affiliation{\chalmersMC}

\date{\today}

\title{Nature and strength of bonding in a crystal of semiconducting nanotubes: 
van der Waals density functional calculations and analytical results}
\begin{abstract}
The dispersive interaction between nanotubes is investigated through 
ab initio theory calculations and in an analytical approximation. 
A van der Waals density functional (vdW-DF) [Phys.~Rev.~Lett.~\textbf{92}, 
246401 (2004)] is used to determine and compare the binding of
a pair of nanotubes as well as in a nanotube crystal. To analyze the 
interaction and determine the importance of morphology, we furthermore
compare results of our ab initio calculations with a simple analytical 
result that we obtain for a pair of well-separated nanotubes.  
In contrast to traditional density functional theory calculations, 
the vdW-DF study predicts an intertube vdW bonding with a 
strength that is consistent with recent observations for the 
interlayer binding in graphitics. It also produce a nanotube wall-to-wall
separation which is in very good agreement with experiments. Moreover, we find that
the vdW-DF result for the nanotube-crystal binding energy can be approximated 
by a sum of nanotube-pair interactions when these are calculated in vdW-DF. This 
observation suggests a framework for an efficient implementation of quantum-physical 
modeling of the CNT bundling in more general nanotube bundles, including nanotube yarn 
and rope structures.

\end{abstract}

\maketitle

\hyphenation{nano-tube}
\hyphenation{nano-tubes}

\section{Introduction}
Carbon nanotubes (CNTs) have a wealth of exciting physical
properties that have made them the focus for a very broad
range of fundamental-science studies.\cite{Dresselhaus} The
CNTs have, for example, an exceptionally large Youngs
modulus.\cite{CNTyoung} The individual CNTs have nanoscale
diameters and micronscale lengths but a range of CNT assembly
processes promise technology applications
even on more macroscopic scales.  Thermal treatment can cause
a fullerene source to transform into a highly regular CNT
crystal with parallel tubes aligned in a
hexagonal structure.\cite{Schlittler} The tubes can also
form CNT bundles\cite{Iijima,CNTstructZhang,Terrones,CNTstructKiang}
in which essentially parallel CNTs still have a very high
degree of (local) order.  The bundles can be spun into
yarn~\cite{LiScience,ZhangScience} and further twisted to
produce torque-free ropes of micrometer diameter (and arbitrary
length). The yarn and ropes have a large strength
and a unique ability to absorb elastic energy in reversible
extensions.\cite{ZhangScience} By pre\-selecting
the nanotube source material, for example, as single-walled
CNTs,\cite{Ericson} it is possible to ensure specific physical
properties (like metallic conductivity) also of the resulting
well-aligned yarn.\cite{ZhangScience,Andrew}

The 
science\cite{Dresselhaus,CNTyoung,Schlittler,Iijima,CNTstructZhang,Terrones,CNTstructKiang} and technology 
progress\cite{LiScience,ZhangScience,Ericson,Andrew} 
challenges us to
present a quantum-physical characterization of the bonding in the
nanotube crystal\cite{Schlittler} and, by extension, in
the bundles. It is valuable to have a method for parameter-free
characterization of general CNT bundles and it is important
to test the accuracy of available computational tools.
The CNT crystals (and bundles) are (approximately) periodic and
have a relatively simple order.  This makes them accessible to
calculations in density functional theory (DFT) which, in principle,
provides quantum-physical accounts of general material bonding.
It is straightforward to provide quantum-physical calculations
for a parameter-free characterization of the intra-CNT electronic
and atomic organization using traditional implementations of
state-of-the-art DFT calculations.\cite{LDAforCNT,GGAforCNT}
These calculations use either the local density approximation
(LDA)~\cite{LDAref} and/or the semilocal generalized gradient
approximation (GGA), for example, as parameterized in the
Perdew-Burke-Ernzerhof flavor.\cite{GGAref} However, the CNT
crystals and bundles are graphitic materials and the intertube
attraction is known to be dominated by relatively soft dispersive or
van der Waals (vdW) interactions.\cite{Schlittler,Kis}
Neither LDA nor GGA provide any physics-based account of the 
bonding between the 
CNTs.\cite{PEexLDAtrouble,Becke,HardNumbers,LayerPRL,Kintercal}

In this paper we use a recently developed van der Waals density
functional\cite{vdWGG,FullggvdW} (vdW-DF) to provide a quantum physical
account of the vdW bonding in a hexagonal crystal\cite{Schlittler} of
parallel semiconducting nanotubes.  We perform state-of-the-art DFT
calculations of the intra-nanotube structure within GGA and of the
inter-nanotube binding within vdW-DF.  The study testifies to the 
strength of this vdW bonding, which is normally described as soft but 
nevertheless contributes significantly to the cohesion of the CNT 
crystal.  Our results allow a test of the vdW-DF theory method by 
comparison against structure measurements for the highly ordered CNT 
crystal\cite{Schlittler} and bundle\cite{Terrones} structures. 
The study supplements a recent vdW-DF calculation\cite{PEcrystal} 
on a simple polymer, polyethylen, for which there also exists experimental
characterization of the crystaline structure.\cite{PEcrystExp}
It also supplements vdW-DF calculations of the benzene and DNA
base-pair interactions\cite{DNAjacs} in a wider program on 
calculating dispersive interactions in carbon and organics 
materials.  We provide a parameter-free theory determination of 
the CNT bonding in the crystal and between pairs of parallel 
nanotubes and document that a summation of nanotube-pair interaction
energies (calculated in vdW-DF) represents a fair approximation for
the vdW-DF results for the crystal bundling energy.  We also detail the 
nature of the mutual CNT interactions by identifying a set of distinct 
vdW interaction regimes.  We show that the vdW interaction is 
significantly enhanced at the bonding separation compared
with the value estimated from the asymptotic interaction.

The vdW-DF calculations correct the accuracy problems arising in
traditional state-of-the-art implementions of DFT (that uses LDA or
GGA) without loss of the traditional-DFT scaling\cite{FullggvdW}
(computation cost increasing $\sim \mathcal{O}(N^3)$ with system size).
DFT calculations in GGA show no meaningful binding.\cite{GGAnobundling}
While LDA calculations can mimic the CNT binding
it underestimates the binding separation. For the CNT bundles,
the LDA result\cite{LDAmimicBundling} for the wall-to-wall
separation, $\Delta_\mathrm{LDA}=\hbox{3.1\ \AA}$, is 10\%
shorter than the experimental value, $\Delta_\mathrm{CNT}
=\hbox{3.4\ \AA}$. Moreover, the LDA result,\cite{LDAmimicBundling}
$\approx 10$ meV/atom, for the intertube binding in a crystal of
metallic (6,6) CNTs is significantly smaller than the
estimate\cite{PAHgraph} (50\ meV/atom for graphitics materials)
extracted from measurements of the binding of polyaromatic hydrocarbon
(PAH) molecules on graphite. The vdW-DF method corrects those problems
without loss of scaling advantages by supplementing the LDA for
correlation with a nonlocal contribution\cite{vdWGG} that scales
like $\mathcal{O}(N^2)$ with the system size.\cite{FullggvdW} 
The vdW-DF method clearly has a better scaling than 
implementations of canonical M{\"o}ller-Plesset
perturbation theory (MP2) for extended structures like polymer
crystals.\cite{TradMP2} Specially adapted MP2 implementations can
achieve a linear scaling with size for large 
molecules.\cite{LinScaleMP2} The adapted MP2
method has also been applied to extended one-dimensional
systems.\cite{Ayala} It is unclear how the adapted-MP2
evaluation of correlation and the vdW-DF determination of
nonlocal correlation compare in actual computing cost
for extended structures like polymer crystals\cite{PEcrystal,Ayala}
and for large bulk and surface-adhesion
systems.\cite{Kintercal,PAHgraphTheory,phenolvdW,SiCadhesion}
In any case, the complete MP2 calculation\cite{Ayala} also 
involves Hartree-Fock calculations which scale worse than 
general DFT implementations (including, for example, vdW-DF).

The paper is organized as follows. In section II we
identify regimes of interactions in crystals of
nanotubes and discuss qualitative differences in the
vdW bonding of semiconducting and of metallic nanotubes.
Section III presents a summary of the vdW-DF calculation
method that we use to obtain an ab initio characterization
of the semiconducting-nanotube crystal.
In section IV we present an analytical description of the nanotube
interaction and in section V we discuss both the strength and
nature of the inter-nanotube interaction.
Section VI contains our conclusions and acknowledgements.

\section{Regimes of nanotube interactions}
\begin{figure}[t]
\begin{center}
\rotatebox{90}{\includegraphics[height=0.95\columnwidth]{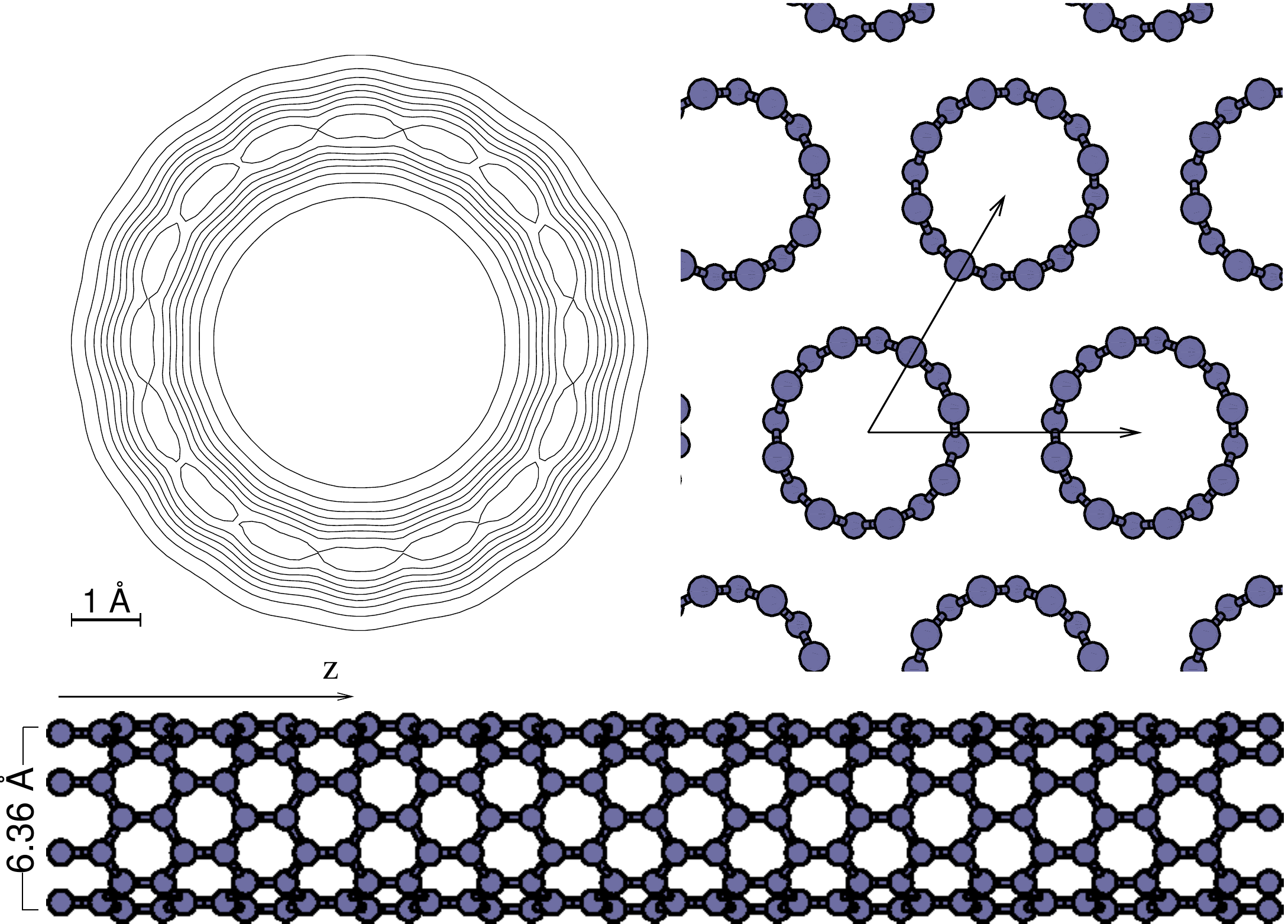}}
\caption{Atomic and electronic structure as well as filament
organization in crystals of semiconducting (8,0) zigzag nanotubes.
\textit{The lower panel\/} shows the fully relaxed atomic configuration
of the individual (8,0) nanotubes as calculated in a traditional
implementation of ab initio density functional theory (DFT).
\textit{The top left panel\/} shows our corresponding traditional DFT
results for the length-averaged electron density concentrated at the
atomic nuclei; the contour spacing is specified in steps of 0.15 e\AA$^3$.
Finally, the \textit{top right panel\/} shows the hexagonal crystalline
order of the nanotube bundle. We calculate the intertube
dispersive interaction and determine the nanotube
crystaline structure using a recently developed
ab initio van der Waals density functional approach.
\protect\cite{vdWGG}
\label{fig:Strucs}}
\end{center}
\end{figure}

There are several regimes of interactions relevant for the
cohesion of the nanotube crystals. The individual tubes are
held together by exceptionally strong covalent bonds between
neighboring carbon atoms separated by just 1.42 \AA.
The binding \textit{between} the CNTs is instead dominated by
the vdW interaction that binds the tubes at a wall-to-wall CNT
separation of 3.4 {\AA}. The vdW interaction also causes an
inter-tube attraction even at asymptotic distances. There are
qualitative differences in the vdW forces in the asymptotic
regime (where the interaction is defined by the dipolar
electrodynamical response) and at bonding separation in
graphitics materials (where multipolar contributions are
documented to enhance the interaction.)\cite{LayerPRL}

Moreover, the vdW interaction between extended semiconducting and
metallic structures (for example, semiconducting and metallic CNTs)
is qualitatively different, at least in the regime of asymptotic
interactions.\cite{Barash,Sernelius,Bostrom,DobsonWhite,FullggvdW}
Single-walled CNTs can exhibit both a semiconducting and
metallic nature of conduction depending on their chirality. Being
low-dimensional systems, the semiconducting and metallic CNTs
therefore exhibit significant differences in their electronic
response and, consequently, in the asymptotic vdW
interactions.\cite{Barash,Sernelius,Bostrom,DobsonWhite}
In the strictly asymptotic regime it is possible to view the CNT as wires.
For a pair of insulating or semiconducting wires, the asymptotic form of
the interaction is known to {\it eventually\/} aquire a $d^{-5}$ dependence
with the separation $d$ between the wire centers of mass. In contrast,
Dobson \textit{et al.} recently used a coupled-plasmon expansion and
approximations valid for asymptotic mutual separations to derive a mutual
interaction energy with a $-d^{-2}(\log (d))^{-3/2}$ asymptotic
scaling for metallic wires.\cite{DobsonWhite}  It is not known to what
extent qualitative differences between the vdW binding of metallic and
semiconducting nanotubes persist down to distances relevant for their
binding in CNT pairs or bundles\cite{FullggvdW} but that is an important
question beyond the present scope. Extraction of CNT binding energies from
the metallic-wire study Ref.~\onlinecite{DobsonWhite} is complicated
because there are two convoluted interaction effects arising as the CNTs
approach each other. Firstly, the CNT morphology (a hollow cylinder)
manifests itself~\cite{SurfSciNT,EMRS} even when the long-wavelength
form of the mutual dielectric response remains applicable.  Secondly,
the nature of the van der Waals interaction changes~\cite{LayerPRL}
so that it is no longer dominated by the long-wavelength response
form, but also retains interaction contributions
defined by the multipole response.\cite{LayerPRL}

Figure \ref{fig:Strucs} shows schematics of the electronic, (intra-tube) 
atomic, as well as (inter-tube) crystalline ordering (bundling) of
nanotubes. We study the mutual binding of pairs and bundles
of CNTs that have a chirality vector\cite{Dresselhaus} (8,0).
These CNTs have a diameter a little larger than 0.6 nm, a
four-fold rotational symmetry, and an along-axis structure
repeating itself every 32 atoms.
Confirming also previous investigations,\cite{GGAforCNT} we find
that state-of-the-art DFT calculations using GGA provide an excellent
account of the intra-nanotube structural organization and electronic
properties such as the nature of conduction.  We use the
traditional DFT-GGA results, obtained in a plane-wave
implementation,\cite{dacapo} as the starting point
for vdW-DF calculations of the intertube binding.\cite{vdWGG,FullggvdW}

Recent density-functional approximations by
us~\cite{LayerPRB,LayerPRL,ijqc,vdWGG,FullggvdW}
and by others~\cite{KohnvdW,Kurth,Dobson,Pitarke,OthervdWDFapplications} 
extend traditional DFT to provide a seamless, parameter-free 
characterization of the vdW
binding without introducing double counting at separations with finite
overlap of electron densities.  In our vdW-DF method\cite{vdWGG,FullggvdW}
we extract the exchange from GGA calculations but supplement the local
density approximation for the correlation energy by a nonlocal correlation
energy contribution $E_\mathrm{c}^\mathrm{nl}$. This contribution is
evaluated from the electron densities of the underlying traditional
DFT calculations in GGA.  This vdW-DF description remains applicable
and effective even for large extended systems that are accessible
for standard ab initio DFT calculation (although at an increase in
computing cost).  In fact, the vdW-DF method exhibits the same scaling
as the underlying traditional DFT calculations.\cite{FullggvdW} The
vdW-DF approach permits ab initio characterizations of large bulk
systems, for example, produced by potassium intercalation.\cite{Kintercal}
It further permits \textit{ab initio} investigations of very large
surface-adhesion systems, for example, graphite-adsorption of
PAH molecules.\cite{PAHgraphTheory,phenolvdW} In a controlled
approximation it even permits an ab initio study (in a repeated
unit cell containing 146 atoms) of the adhesion of graphite
sheets on SiC surfaces.\cite{SiCadhesion}

The vdW-DF evaluation of the nonlocal correlations $E_\mathrm{c}^\mathrm{nl}$
(vdW interaction energy) involves a density-weighted integration of a
kernel~\cite{vdWGG} that contains a rich account
of the complex electrodynamics.\cite{LayerPRL,ijqc,vdWGG,FullggvdW}
Our vdW-DF is not developed to include an explicit
account of the asymptotic interaction between extended
metallic one-dimensional systems.\cite{FullggvdW}
The form of the vdW-DF kernel\cite{vdWGG,FullggvdW} ensures
the correct asymptotic behavior of vdW interactions for
atoms, molecules, and most surface and bulk systems. It also
describes the asymptotic interactions for extended
low-dimensional systems that are isolating or semiconducting.
The form of the vdW-DF kernel ensures the correct asymtotic
form of the interaction between pairs and within crystals of the
(8,0) CNTs because these are robustly semiconducting (characterized
by a significant gap~\cite{Dresselhaus}). More importantly,
our vdW-DF calculations of the (8,0)-CNT binding not only
{\it eventually\/} reproduces a $d^{-5}$ dependence in the
interaction energies but reveals a much finer structure and
remains fully applicable at general separations.

To interpret this rich structure in CNT binding-energy variation with
CNT separation, we also present in this paper an analytical evaluation of
the CNT-pair interaction, valid at intermediate-to-asymptotic distances.
Our analysis tool (but not our full vdW-DF calculations) makes
assumptions of non-overlapping electron densities and of a
long-wavelength form of the CNT dielectric response but it
respects the CNT morphology.\cite{EMRS,OtherFormal,MSEvdWNT}
Comparison with the full vdW-DF calculations therefore allows us to
deconvolute effects arising from the change in nature of the
dielectric response.  We thereby provide an analysis
that splits the interaction into two major regimes: (1)  a close regime at or
near binding separations where full ab initio vdW-DF calculations are essential
for an accurate account and (2)  an intermediate-to-asymptotic regime where the
long-wavelength dielectric response remains applicable but
where the CNT morphology specifies the variation of the interaction.
The result of this analys documents that the vdW interaction enhances
at bonding separations compared with estimates that can be extracted
from knowledge of the asymptotic form of the inter-nanotube bonding.

\section{Computational Methods}

The top right panel of Fig.~\ref{fig:Strucs} shows a schematic
of the repeated two-dimensional hexagonal array of the nanotube bundles.
We apply the vdW-DF~\cite{vdWGG} method to include the dispersive 
interaction within the framework of traditional plane-wave DFT both 
for dimers of nanotubes as well as for an infinitely extended 
nanotube crystal (Figure \ref{fig:Strucs}).  A self-consistent formulation of vdW-DF has 
recently been derived, implemented, and tested.\cite{FullggvdW} 
Here we use the original, non-selfconsistent (post-GGA) 
implementation~\cite{vdWGG} that rests upon and utilizes 
traditional DFT calculations to obtain the electron density variations. 

In the present vdW-DF study we furthermore take advantage
of the success of the traditional semi-local (GGA) density 
functionals to describe the {\it{intra}}-molecular properties 
of the nanotubes (as well as the electron densities).  It is,
in principle, possible to provide an all-vdW-DF characterization
of the intra-nanotube atomic structure (allowing relaxation under
vdW-DF forces\cite{FullggvdW}) but the computation costs
would be large. Our previous experience from an all-vdW-DF
characterization of a single graphite sheet~\cite{Kintercal} 
suggests that only minute differences would result for the
CNT structure if we replaced the GGA intra-tube characterization
by a full vdW-DF characterization.

\subsection{Nanotube atomic and electronic structure}

\begin{figure}[t]
\begin{center}
\includegraphics[width=0.80\columnwidth]{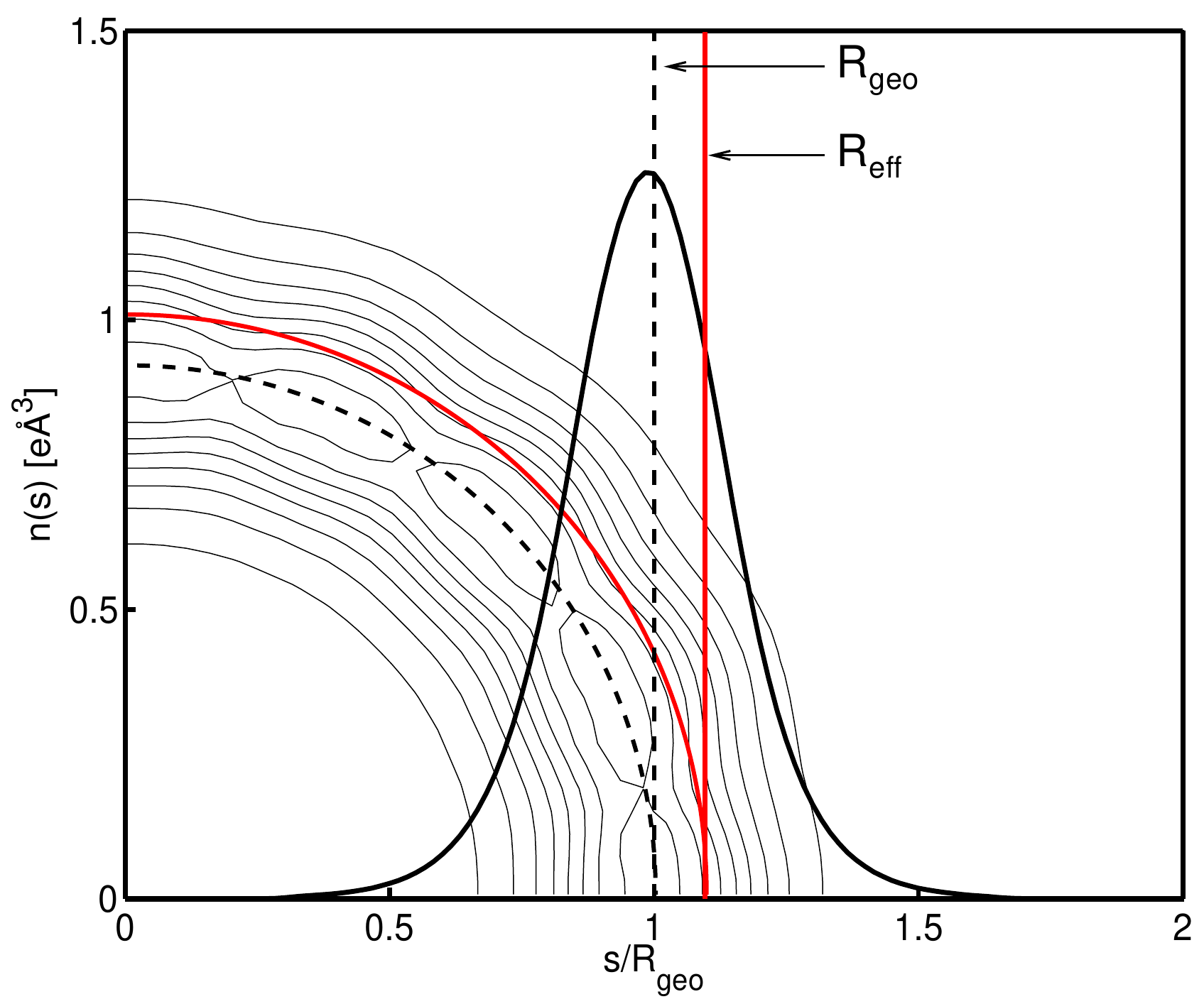}
\caption{\label{fig:response}Length- and radially-averaged
electron density $\bar n(s)$ shown together with the positions
of the effective and the geometric radius.
The radial separation is given relative to the geometric
radius of the nanotube. 
The background inset shows the length averaged electron
density with contour lines separated in steps of 0.15 e\AA$^3$.}
\label{fig:radius}
\end{center}
\end{figure}

In the actual vdW-DF implementation, a large set of state-of-the-art 
traditional DFT calculations determine the atomic structure
(of the individual nanotube) and the electron density
variation of the nanotubes when isolated, when assembled into a 
hexagonal crystaline structure, or when aligned as a parallel nanotube 
dimer.  We use a plane-wave code~\cite{dacapo} with ultra-soft 
pseudopotentials and a $1\times 1\times 8$ 
Monkhorst-Pack~\cite{Monkhorst1972} $k$-point 
sampling of the Brillouin zone of the 
periodically repeated unit cell containing 32 atoms 
per nanotube.  We perform self-consistent
calculations in the Perdew-Burke-Ernzerhof (PBE)-flavor\cite{GGAref} 
of GGA for the exchange and correlation functional.  We choose a plane-wave 
energy cut-off 476 eV and specify the Fast Fourier Transform (FFT) grid so 
that the (density) grid spacing remains smaller than 0.14 {\AA} in all 
directions.  

We first determine the atomic structure of the individual (8,0)
nanotube, lower panel of Fig.~\ref{fig:Strucs}.  We use our ab initio 
calculations of the strong intra-nanotube atomic forces to relax the 
morphology to a total residual atomic force 0.05 eV/{\AA} per unit 
cell.  The structure is characterized by an effective geometrical radius
$R_\mathrm{geo}$ specified as the average distance of the carbon 
nuclei from the axis defined by the nanotube center of mass.  We 
determine the value of that geometric radius to be 
$R_\mathrm{geo}=\hbox{3.18 \AA}$.  As explained above, the atomic 
structure (obtained in PBE-GGA for the isolated nanotube) is kept 
frozen in all subsequent calculations (for nanotube crystal and 
dimer cases). We have explicitly tested that no additional intratube
atomic-relaxation is relevant at (or beyond) the separation that
characterizes the nanotube-crystal/dimer binding in
our subsequent post-GGA vdW-DF characterization.

Next we determine the electronic structure from traditional DFT 
calculations for the isolated nanotubes, for the nanotube bundles,
and for the dimer structures. The upper left panel of Fig.~\ref{fig:Strucs}
shows contours (at 0.15 e\AA$^{-3}$ interval) in the isolated-nanotube
electron density variation averaged along the nanotube axis. The maximum 
in electron density coincides with the radial position of nuclei and
naturally identifies the nanotube wall. To characterize the inter-nanotube 
vdW-dominated binding we further calculate the electron density for 
nanotubes in hexagonal crystals and dimers as a function of the 
wall-to-wall nanotube separation $\Delta$ and as a function of the 
nanotube rotation angle (relative rotation angle in the case of the 
dimer study).
 
We find that there are important electron-density overlaps for
$\Delta < \hbox{4 \AA}$ when the self-consistent GGA
electron densities differ from a superposition of 
individual-nanotube electron densities. The existence
of these intertube electron-density overlaps has direct
consequences for the details of the intertube vdW binding.
However, the electron-density overlap does not reflect the 
existence of any relevant (and physically meaningful) binding 
arising within the GGA calculations themselves (in any of the
GGA flavors). 

\subsection{van der Waals density functional theory}

Like the LDA and GGA functionals, the vdW-DF is defined by 
approximations for the exchange and correlation energies.
The nonlocal dispersive interactions responsible for keeping the 
bundle of nanotubes together are included as a significant 
extension (correction) of the correlations energy in the 
underlying GGA calculations.  Specifically, we completely replace 
the GGA description of correlation but use the self-consistent GGA 
result for the electron density to evaluate a new correlation energy that
includes the nonlocal nature of the vdW binding.

Our vdW-DF splits the correlation up into local and nonlocal 
contributions:\cite{LayerPRB,vdWGG}
\begin{equation}
E_\mathrm{c}
\approx
E_\mathrm{c}^{\mathrm{LDA}}+E_\mathrm{c}^\mathrm{nl},
\end{equation}
with the local part approximated in the LDA.  The nonlocal 
correlation energy is expressed~\cite{LayerPRB,vdWGG}
\begin{equation}
E_\mathrm{c}^\mathrm{nl} = \int_0^\infty \frac{du}{2\pi}
\mathrm{tr}[\ln(1-V\tilde{\chi})-\ln(\epsilon)]
\label{eq:FullEcnl}
\end{equation}
where $u$ is the imaginary frequency, $V$ is the interelectron 
Coulomb interaction potential, $\tilde{\chi}$ is the local-field 
density response. The isotropic dielectric function,
$\epsilon=\mathrm{tr}(1+4\pi\mathbf{\alpha}))/3$, is also specified by 
the local-field density response, $\tilde{\chi}=\nabla\cdot
\mathbf{\alpha}\cdot\nabla$.  The nonlocal correlation energy 
is further approximated 
\begin{equation}
E_\mathrm{c}^\mathrm{nl}[n]=\frac{1}{2}
\int d\vec r\,d\vec r' n(\vec r)\phi(\vec r,\vec r')n(\vec r'),
\label{eq:kernel}
\end{equation}
through a kernel $\phi$ specified by a number of sum 
rules~\cite{vdWGG} and physics results.\cite{vdWGG,FullggvdW,Vosko}
The interaction energy~(\ref{eq:kernel}) is consistently constructed 
to vanish for a homogeneous system. The kernel $\phi$ is specified by a pair of local parameters
$q_0(\vec r)$ and $q_0(\vec r')$ that depend on the electron 
density and the density gradient.  The kernel can  be tabulated 
in advance in terms of an effective separation 
$D=[(q_0+q_0')/2]|\vec r-\vec r'|$  and an asymmetry 
parameter $\delta=(q_0-q_0')/(q_0+q_0')$.  The value of the 
$q_0$'s are chosen to reproduce the (plasmon-pole) response of a weakly 
perturbed electron gas, Ref.~\onlinecite{vdWGG}.

With the  evaluation of the nonlocal energy contribution 
(from underlying GGA calculations of the electron densities)
we arrive at a vdW-DF total energy calculation:\cite{ijqc}
\begin{eqnarray}
E^\mathrm{vdW-DF}
& = & E^0+E_\mathrm{c}^\mathrm{nl},
\label{eq:EvdWDF}\\
E^0 & = & E^\mathrm{GGA}-E^\mathrm{GGA}_\mathrm{c}+E^\mathrm{LDA}_\mathrm{c}.
\label{eq:Eloc}
\end{eqnarray}
The GGA energy term $E^\mathrm{GGA}$ is here evaluated in 
the revPBE~\cite{GGArevPBE} flavor (based on the self-consistent
calculations for the electron density that we obtain in the PBE
flavor of GGA).  Effectively this amounts to a small adjustment of
the exchange contribution, which we do to minimize any potential
artificial exchange binding in the plane-wave formalism 
used.\cite{LayerPRL,ijqc,revPBEWu,vdWGG}
The new (semilocal) energy contribution $E^0$ represents a 
modification of GGA that, for example, retains a description of 
the kinetic-energy repulsion as well as, for example, 
covalent\cite{phenolvdW} or ionic\cite{Kintercal} interactions.

\subsection{van der Waals density functional calculations}

The evaluation of the nonlocal correlation $E_\mathrm{c}^\mathrm{nl}$ 
requires extra care due to a grid sensitivity of the functional 
form (\ref{eq:kernel}).  The vdW binding in the nanotube crystal
and dimer cases arises almost exclusively from a difference
in nonlocal-correlation energy $E_\mathrm{c}^\mathrm{nl}$ for the 
crystaline/dimer-structure and for the isolated nanotubes.
However, the intramolecular density variation causes a 
very large contribution to $E_\mathrm{c}^\mathrm{nl}$ that 
must be carefully subtracted in our ab initio calculations of the 
inter-nanotube interaction.  Moreover, the evaluation of this 
intra-nanotube $E_\mathrm{c}^\mathrm{nl}$ energy is somewhat
sensitive to the relative position of atomic positions and FFT 
grid points.\cite{Kintercal}  Nevertheless,
robust and efficient evaluation of the inter-nanotube binding is 
possible by simply ensuring that we calculate and subtract 
$E_\mathrm{c}^\mathrm{nl}$ contributions for the isolated 
nanotube using FFT grid points that closely match those of 
the composite system, as further explained in Ref.~\onlinecite{Kintercal}.

In practice we evaluate the vdW-DF binding in the nanotube crystals
and dimers by supplementing every interacting nanotube system 
by a suitable reference calculation of the isolated nanotube.
We determine the binding in nanotube bundles by adding (at every 
nanotube separation) a DFT calculation of the electron density for 
a corresponding isolated nanotube located in a cell of double size 
in each of the two perpendicular directions and on a FFT grid that 
retains the absolute grid-point spacing.\cite{Kintercal}
In our calculations of $E_\mathrm{c}^\mathrm{nl}$, for the underlying 
GGA calculation we use a FFT-grid spacing that is always smaller 
than 0.14 \AA\ in any direction; this choice is found 
sufficient given our work to carefully synchronize the FFT gridding 
when we calculate the electron density for the interacting and isolated 
nanotube system.

The real-space implementation of $E_\mathrm{c}^\mathrm{nl}$ is
simply applied to extended systems as graphite~\cite{Kintercal} and
polyethylene\cite{PEcrystal} by evaluating 
$E_\mathrm{c}^\mathrm{nl}$ for the unit cell electron density 
as well as  the nonlocal interaction from its surrounding images. 
To accelerate the vdW-DF characterization
we limit the evaluation of the multidimensional integral 
(\ref{eq:kernel}) to contributions from grid points having
a density larger than 10$^{-4}$ a.u.  The use of such a 
density cut-off is strongly motivated by the excellent convergence that 
we have previously documented for graphitic systems even when a significant ionic bond 
supplements the binding from nonlocal correlations.\cite{Kintercal}
The nonlocal correction from the surrounding electron density 
converges rapidly in terms of the separation to the unit cell. 
We have tested that it is in general sufficient to only 
include the interaction from the electron density that is less than 
12 \AA\ away from the unit-cell boundaries in the direction along the 
nanotube and 15 \AA\ in the directions perpendicular to the nanotube axis.
Nevertheless, to converge the $E_\mathrm{c}^\mathrm{nl}$ calculations
to a sub-meV level and retain a very high relative accuracy even in 
the asymptotic regime, we choose to retain 
$E_\mathrm{c}^\mathrm{nl}$-contributions 
originating from points closer than 24 \AA\ (and in some cases even
closer than 30 \AA) in the direction of the CNT extension. 

\section{van der Waals interactions at 
intermediate to asymptotic distances}

From our ab initio vdW-DF calculations of the asymptotic van der Waals
interactions we extract an analytical determination of the van 
der Waals interaction energy $E_\mathrm{vdW}$ per unit length $L$
for a nanotube pair as a function of the separation 
$d=2R_\mathrm{geo}+\Delta$. The analytical result for 
$E_\mathrm{vdW}(d)/L$ rests on the approximation summarized in 
Refs.~\onlinecite{SurfSciNT,MSEvdWNT,EMRS}.  It assumes that the 
electron densities of the two nanotubes do not overlap and 
constitute a lowest-order expansion~\cite{BILijqc} of 
Eq.~(\ref{eq:FullEcnl}) in terms of the {\it external-field\/} 
susceptibilities, $\mathrm{\alpha}_\mathrm{eff}$: 
\begin{equation}
E_\mathrm{vdW}=-\int_0^\infty \frac{du}{2\pi}
i\mathrm{tr}[\alpha_\mathrm{eff,1}T_{12}\alpha_\mathrm{eff,2}T_{21}].
\label{eq:EvdWapprox}
\end{equation}
Here $T_{ij}$ denotes the dipole interaction tensor,
$T_{ij} = - \nabla_i\nabla_j|\vec r_i-\vec r_j|^{-1}$.
The analysis is possible to carry out for nonisotropic external-field 
susceptibilities,\cite{EMRS,SurfSciNT} but for an interpretation of our 
vdW-DF calculations of CNT interactions it is sufficient to consider 
isotropic  susceptibilities $\alpha_\mathrm{eff}$. We focus on the 
interaction regime where effects of the CNT morphology dominates 
the variation of $E_\mathrm{c}^\mathrm{nl} \approx 
E_\mathrm{vdW}$ with distance.\cite{SurfSciNT,MSEvdWNT} We assume a 
long-wavelength form of $\alpha_\mathrm{eff}$ so that the resulting analytical 
determination remains valid at such intermediate-to-asymptotic interaction 
distances.

The physics of the local-field and external-field susceptibilities
defines the parameterization of our vdW-DF 
method.\cite{LayerPRB,LayerPRL,vdWGG,FullggvdW}
The long-wavelength electrodynamical response determines the interaction at large
distances\cite{ijqc} and our vdW-DF method describes this response by the 
isotropic effective (external-field) susceptibility~\cite{vdWGG}
\begin{equation}
\alpha^\mathrm{gg}_\mathrm{eff}(u;\vec r)=\frac{n(\vec r)}
{u^2+[9q^2_0(\vec r)/(8\pi)]^2}.
\label{eq:effsup}
\end{equation}
We stress that our extraction of this long-wavelength form serves only 
to establish formal connection between the full vdW-DF calculations and 
the analytical approximation.  We also emphasize that neither the effective 
response (\ref{eq:effsup}) nor the full vdW-DF response function~\cite{vdWGG} 
is explicitly designed to accurately reproduce, for example, the static 
dielectric response, in contrast to the functional approaches 
described in Refs.~\onlinecite{LayerPRL,ijqc,EMRS} and \onlinecite{SurfSciNT}. 
Rather, the full vdW-DF response function~\cite{vdWGG} is constructed exclusively 
from an ansatz for the plasmon-pole response, conservation rules and many-body 
calculations~\cite{vdWGG,Vosko} to describe the average response. The full
vdW-DF description involves contributions from different frequencies and 
wavelengths and it is the average response, rather than the long-wavelength 
limit (\ref{eq:effsup}), that determines the interactions at binding 
separations where our vdW-DF approach is most needed.

Our analysis focus on the intermediate-to-asymptotic separations furthermore
allows us to consider the contributions to the susceptibility from the
electron density averaged over the angular and
along-tube variations. We thus substitute
$\alpha_\mathrm{eff}(\vec r;u)
\rightarrow \bar \alpha_\mathrm{eff}(s;u)$, where $s$
denotes the radial distance from the nanotube center.
The nanotube interaction per unit length, given in terms of the 
effective response (\ref{eq:effsup}), becomes:\cite{EMRS}
\begin{align}
\frac{E_\mathrm{vdW}}{L}=&-\int_0^\infty  \frac{du}{2\pi}\int_0^\infty 
\,ds_1 s_1
\int_0^\infty 
\,ds_2 s_2\label{eq:semiasymint}\\
&\quad\bar \alpha_\mathrm{eff}(s_1;u)\bar\alpha_\mathrm{eff}(s_2;u)
\nonumber
\!\!\!\!\sum_{\alpha,\beta=s,\theta,z}\!\!\!\!G_{\alpha,\beta}(s_1,s_2),
\intertext{with the geometry factors}
G_{\alpha\beta}=&\int_0^{2\pi}d\theta_1  
\int_0^{2\pi} d\theta_2\int_0^\infty d(z_2-z_1)\\ 
&[T^{\alpha\beta}_{12}(s_1,\theta_1,z_1;s_2,\theta_2,z_2)]^2,\nonumber
\end{align} 
where the dipole interaction tensors are expressed in cylindrical
coordinates.  The result~(\ref{eq:semiasymint}) is easily 
expanded in the inverse center-of-mass separation $d^{-1}$
of the nanotubes, yielding interaction energies of the form
\begin{equation}
\frac{E_\mathrm{vdW}}{L}=-\frac{B_5}{d^5}-\frac{B_7}{d^7}+\ldots,
\label{eq:expansioncoff}
\end{equation}
with $B_5$ and $B_7$ given by
\begin{align}
B_5&=\frac{9}{8}\int_0^\infty  du\, \Xi^{(0)}(u)^2,\\
B_7&=\frac{225}{16}\int_0^\infty   du\, \Xi^{(0)}(u)\Xi^{(2)}(u).
\intertext{Here}
\Xi^{(i)}(u)&\equiv \int_0^\infty d s\, 
2\pi s \bar \alpha_\mathrm{eff}^\mathrm{gg}(s;u) s^i,
\end{align}
is simply the $i$'th moment of the effective response. We calculate
the coefficients $B_{5,7}$ directly from the $\alpha_\mathrm{eff}(s;u)$
variation specified by our underlying GGA-DFT calculations of the
CNT electron density variation. We have explicitly tested 
consistency of this asymptotic evaluation and the set of full vdW-DF
calculations for $\Delta > \hbox{16--20 \AA}$.

The relevant external-field electrodynamical response 
($\alpha_\mathrm{eff}$) of the nanotubes is dominated by 
contributions at some radius $R>R_\mathrm{geo}$. That is the
experience gained from describing the electrodynamical response 
and van der Waals interactions of surfaces~\cite{Hult} and from
previous investigation of the van der Waals bonding in 
graphitics.\cite{LayerPRL,Kintercal}  While the results
presented in Refs.~\onlinecite{SurfSciNT,MSEvdWNT,EMRS} made the 
assumption that the response $\alpha_\mathrm{eff}$ arose exclusively from
the atom wall (at $R_\mathrm{geo}$), the formal special-function
evaluation~\cite{EMRS,Glasser} 
\begin{equation}
\frac{E_\mathrm{vdW}(d,R)}{L}=-
\frac{B_5}{d^5}{}_3F_2\left({\textstyle \frac{1}{2}},
{\textstyle \frac{5}{2}},{\textstyle \frac{5}{2}};1,1;\frac{4R^2}{d^2}
\right),
\label{eq:hyperexpress}
\end{equation}
is possible as long as we may assume the response $\alpha_\mathrm{eff}$ 
dominated by contributions at any (single) radius $R$. The interaction
result (\ref{eq:hyperexpress}) simply reflects the morphology (interaction
of two hollow cylinders). The effective vdW-DF response 
$\alpha_\mathrm{eff}^\mathrm{gg}(s;u)$ is dominated by contributions 
at $R_\mathrm{geo}$ and outside in the CNT density tails.
In this paper, we choose a value for the effective response radius
\begin{equation}
R_\mathrm{eff}=\sqrt{\frac{2 B_7}{25 B_5}}
\end{equation}
and obtain an analytical approximation
\begin{equation}
E_\mathrm{c}^\mathrm{nl}(\Delta)
\approx E_\mathrm{vdW}(d=2R_\mathrm{geo}+\Delta,R_\mathrm{eff}).
\label{eq:analyticEcnl}
\end{equation}
that exactly reproduces the asymptotic variation of the
full vdW-DF calculations up to the second spatial moment given by 
$B_5$ and $B_7$.  The relative position of $R_\mathrm{eff}$ and 
$R_\mathrm{geo}$ are shown in Fig.~\ref{fig:radius}.

\section{results and discussion}

\begin{figure}
\begin{center}
\includegraphics[width=0.9\columnwidth]{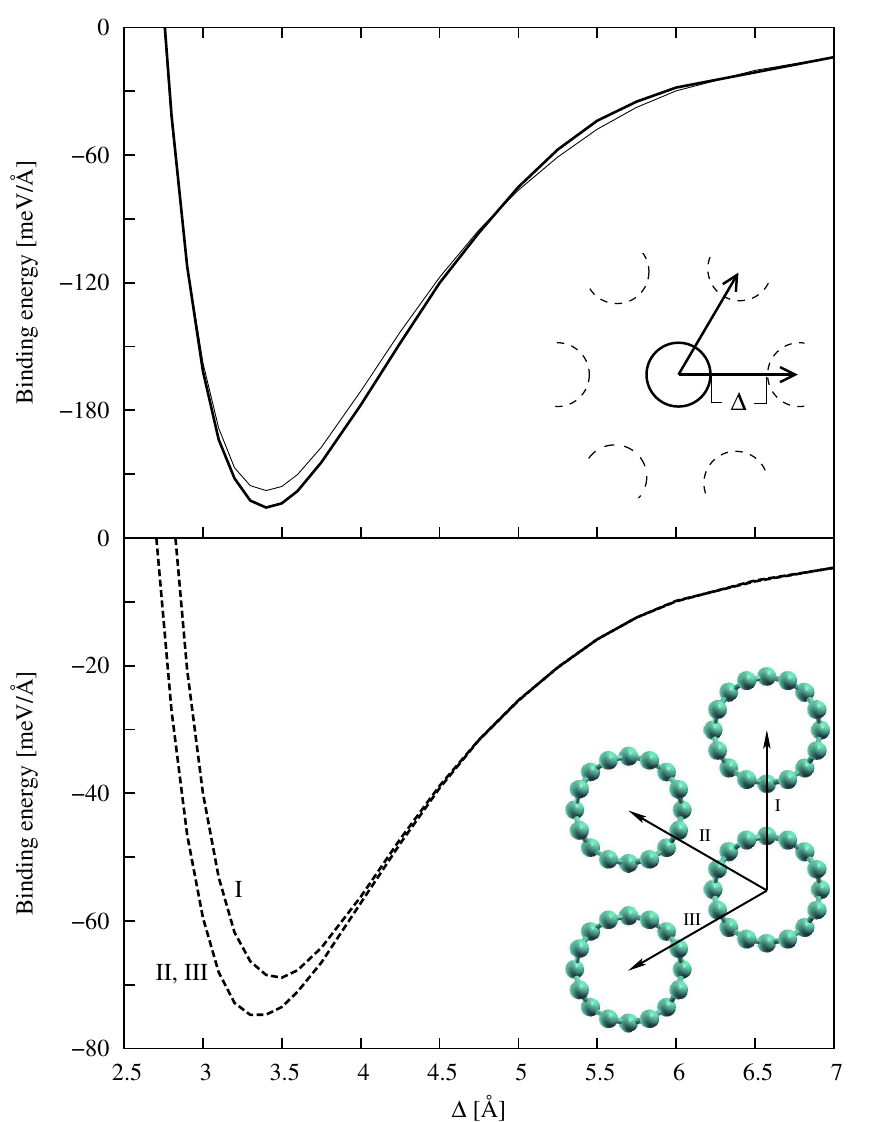}
\caption{\label{fig:BindEng} 
Nanotube binding energy 
$E^\mathrm{vdW-DF}(\Delta)-E^\mathrm{vdW-DF}(\Delta\to\infty)$,
(per unit length)
for semiconducting (8,0) nanotubes evaluated in vdW-DF as a function of the 
wall-to-wall separation $\Delta$.  {\it The top panel\/} shows 
vdW-DF results for the hexagonal crystal and compares the 
crystal interaction energy (thick solid curve) against an estimate 
(thin solid curve) based on a sum of vdW-DF results for the 
nanotube pair interactions.  {\it The bottom panel\/} reports 
vdW-DF calculations of the binding of two parallel nanotubes 
(dashed curves) in three different atomic configurations indicated 
in the insert. The sum of those three pair interactions constitutes
the approximation for the crystal interaction (thin solid curve)
in the top panel.
}
\end{center}
\end{figure}

\subsection{van der Waals bonding in a nanotube crystal}

The top panel of Fig.~\ref{fig:BindEng} reports our ab initio calculation 
(thick solid curve) of the vdW-DF binding in a hexagonal crystal of semiconducting 
(8,0) nanotubes.  The vdW-DF result for the binding separation, $\Delta_\mathrm{bind}
\approx \hbox{3.45 \AA}$, is in very good agreement with experimental
observations,\cite{Iijima,Terrones,CNTstructZhang,CNTstructKiang} 
$\Delta_\mathrm{exp}\approx\hbox{3.4 \AA}$, and corrects the poor 
structure predicted in traditional LDA calculations\cite{LDAmimicBundling}
$\Delta_\mathrm{bind}^\mathrm{LDA}=\hbox{3.1 \AA}$.
The binding energy for the nanotube bundle is
large, $E_\mathrm{bind}^\mathrm{crys}=-30$ meV/atom corresponding to $
-0.225$ eV/\AA, consistent with interaction strengths that 
we have previously calculated in vdW-DF for the interlayer binding in 
graphite:\cite{Kintercal} $-50$ meV/atom.
The vdW-DF binding energy is significantly larger than the LDA result,\cite{LDAmimicBundling}
$\approx 10$ meV/atom, obtained for a metallic (6,6) nanotube.

The figure documents differences between the vdW-DF calculation for the
full CNT crystal (thick solid curve) and corresponding approximations based 
on CNT-pair contributions (thin solid curve).  The regular 
vdW-DF calculations yield a CNT-crystal binding energy that 
is larger than the binding energy estimate obtained from the 
summation of pair contributions $E_\mathrm{bind}^\mathrm{crys,est}
=-29$ meV/atom corresponding to $-0.220$ eV/\AA. 
We find that the vdW-DF energy difference 
$E_\mathrm{bind}^\mathrm{crys}-E_\mathrm{bind}^\mathrm{crys,est}$
is split evenly between contributions $E_\mathrm{c}^\mathrm{nl}$ and $E_0$.

Nevertheless, the vdW-DF results for the pair interaction energies constitute 
a fair approximation of the hexagonal ordering arising in the nanotube
bundles. It is thus possible to use vdW-DF calculations of the CNT-pair
interactions at general (parallel) configurations (of different relative
rotations) to model the cohesion and binding in more general nanotube
structures such as yarn and rope.

\subsection{van der Waals bonding in a pair of parallel semiconducting
nanotubes}

Fig.~\ref{fig:BindEng}, bottom panel, reports our ab initio calculation 
of the vdW bonding between pairs of parallel (8,0) nanotubes at three
configurations `I', `II', and `III', identified in the insert. These
are the configurations that are relevant for the pair-interaction estimate
of the CNT hexagonal crystal (thin solid curve in top panel).
Even for a CNT pair, the nanotube binding is very significant, 
$E_\mathrm{bind}^\mathrm{pair} \approx -9.2$ meV/atom, but occurs
at slightly different binding separations for different relative 
nanotube rotations. We find that the vdW-DF
results for the nonlocal correlation term $E_\mathrm{c}^\mathrm{nl}$
are almost identical (smaller than 1\% variation outside binding
separations) for the three CNT-pair configurations.  As is evident in 
the insert (which identifies actual atomic organization investigated in 
our vdW-DF method,) the atomic organization is in better registry for 
some organization than others. There consequently exists some 
electron-density variation with the rotations and our vdW-DF method 
is sensitive to that variation since the semilocal contribution 
$E_0$ contains a description of the kinetic-energy repulsion. 

As an interesting aside, we note that the high symmetry of the semiconducting 
(8,0) nanotube permits us to test the grid-sensitivity and consistency of the vdW-DF 
calculations. There must exist a four-fold symmetry in atomic positions 
around the (8,0) nanotube and such an approximate symmetry
also emerges as a result of the initial atomic relaxation that 
we perform for an individual nanotube in traditional DFT. The symmetry
implies a periodicity $\pi$ in the variation of the vdW binding between a pair of nanotubes
with the relative rotation angle 
$\Theta$.  
However, the imperfect relaxation causes small variations in the exact atomic 
location relative to the grid.
We find that the vdW-DF calculations are more sensitive
than the underlying traditional DFT calculations.
Nevertheless, the vdW-DF calculations respect the symmetry and
produce vdW interaction energies for $\Theta$ and $\Theta+\pi$ 
relative rotations that are identical even at a sub-meV energy scale.

\subsection{Approximative microsopic modeling for general nanotube-bundle structures}

The comparison between the vdW-DF results for the nanotube crystal and for the
approximation based on a sum of nanotube-pair interactions, Fig.~\ref{fig:BindEng},
top panel, suggests a framework for an approximative microscopic modeling for the 
binding in more general bundles of (semiconducting) nanotubes.
A simple mapping of the binding energy for two parallel 
nanotubes for all combinations of independent rotations (relative to the 
interaction line) provides the starting point. Adding such general 
pair-interaction contributions allows vdW-DF calculations to 
account for general vdW bonding in aligned nanotube structures, 
including nanotube yarn and ropes.

Moreover, the finding of insignificant differences between the $E_\mathrm{c}^\mathrm{nl}$ 
energy contributions for the three nanotube-pair configurations investigated in 
Fig.~\ref{fig:BindEng}, suggests an additional speed up in the modeling.
Assuming that general, independent nanotube rotations also 
causes insignificant $E_\mathrm{c}^\mathrm{nl}$ differences,
it is sufficient to supplement one calculation of $E_\mathrm{c}^\mathrm{nl}$ 
(detailed below) with a mapping of the general $E^\mathrm{0}$ variation. This
can be obtained at a computational cost equal to that of traditional implementations 
of DFT.  A forthcoming study will provide vdW-DF results for the bundling of a broader
set of semiconducting nanotubes and an explicit test of the $E^\mathrm{vdW-DF}$ and 
$E_\mathrm{c}^\mathrm{nl}$ variation with general (independent) nanotube-rotation angles 
to detail the suggested approximative modeling approach.

\subsection{Nature of the vdW bonding at bundle and at intermediate 
separations}

\begin{figure}[t]
\begin{center}
\includegraphics[width=0.85\columnwidth]{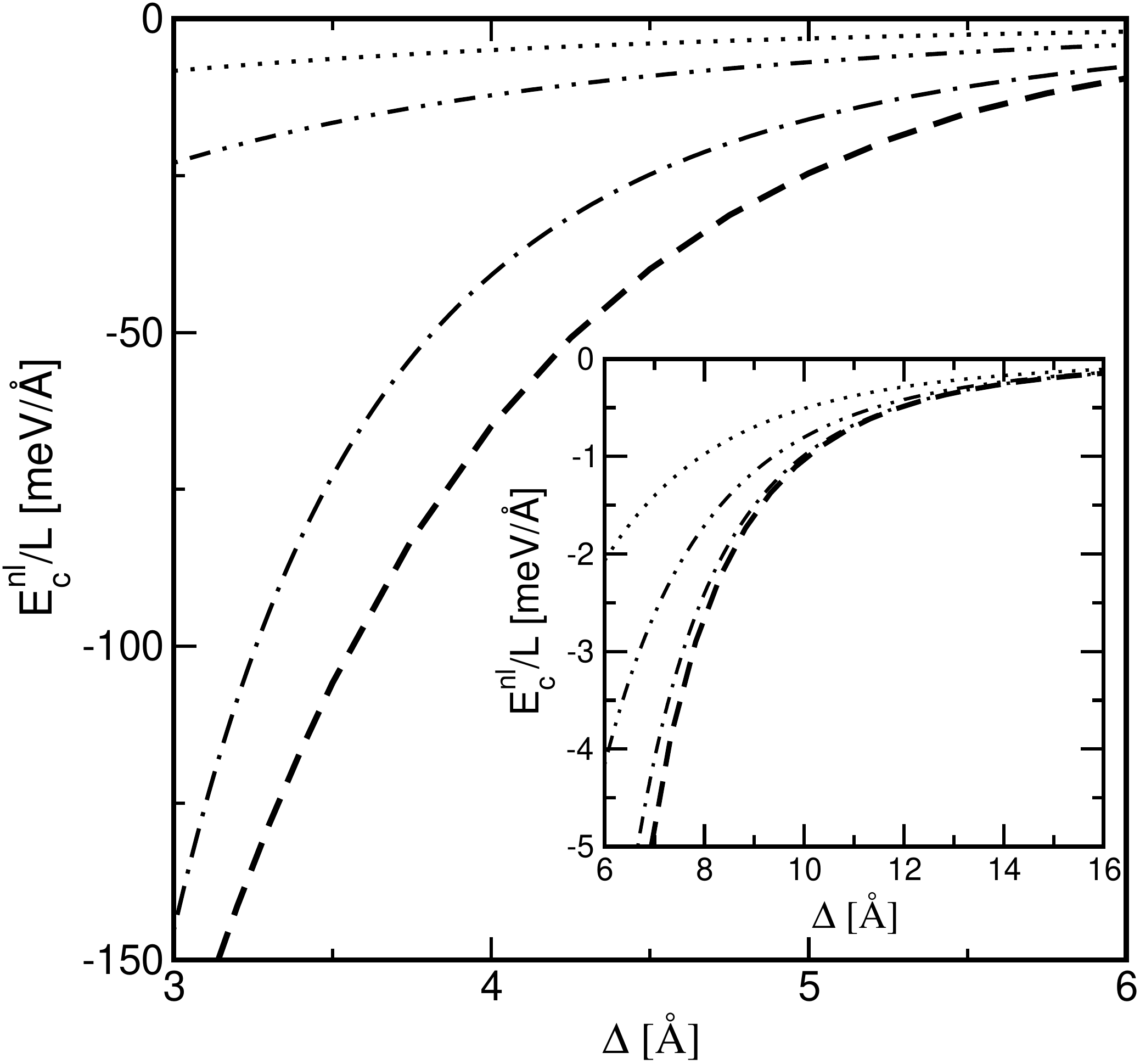}
\caption{\label{fig:hyperfit} Nonlocal correlation energy per unit 
length for a nanotube dimer near binding separations (main panel)
and in the intermediate to asymptotic regime (insert panel).  The thick 
dashed curves show the results of the full vdW-DF calculation of the nonlocal 
correlation energy (vdW interaction).  The dotted and dashed-double-dotted
curves show the (traditional) asymptotic interaction estimates
as determined from the asymptotics  $-B_5d^{-5}$ and
$-B_5d^{-5}-B_7d^{-7}$, respectively. Finally, the
dashed-dotted curve shows the analytical evaluation
that approximates the electrodynamical response by
the long-wavelength form but retains a full description
of the nanotube morphology. The analytical result
also reflects the surface-physics insight that the
electrodynamical response is dominated by contributions
outside the radius defined by the atomic positions, 
Fig.~2.}
\end{center}
\end{figure}

Fig.~\ref{fig:hyperfit} compares the full vdW-DF calculation
of $E_\mathrm{c}^\mathrm{nl}$ contribution to the CNT-pair interaction, 
thick dashed curves, with vdW-DF based approximations $E_\mathrm{vdW}$, 
dotted and dash-dotted curves, near binding separations (main panel) 
and in the intermediate-to-asymptotic regime (insert). The contribution 
$E_\mathrm{c}^\mathrm{nl}$ is evaluated for the configuration 'I' shown 
in the insert of the lower panel of Fig.~\ref{fig:BindEng} (but $E_\mathrm{c}^\mathrm{nl}$
exhibits only insignificant differences between configurations `I', `II' and `III').
All of the estimates $E_\mathrm{vdW}$ are, of course, independent of the
nanotube rotation by construction. The dashed-dotted curves show the 
analytical CNT-pair interaction estimate (\ref{eq:analyticEcnl}) that
invokes a long-wavelength form of the electrodynamical
response but respects the morphology of the interaction 
problem.\cite{SurfSciNT,EMRS}
The dotted and dashed-double-dotted curves show (for 
$d= \Delta+ 2R_\mathrm{geo}$) traditional interaction 
estimates, $-B_5/d^5$ and $-B_5/d^5-B_7/d^7$ respectively.
The traditional interaction estimates clearly only become
applicable in a very remote asymptotic regime beginning at 
$\Delta > \hbox{16 \AA}$.

The main panel shows that the full vdW-DF calculations are necessary
around the binding separations $\Delta_\mathrm{bind} \sim \hbox{3.5 \AA}$.
Here the interaction is significantly enhanced compared with estimates
based on the asymptotic dipolar response.  The enhancement is consistent with 
the behavior documented for graphite interactions as described in an earlier 
generation of vdW-DF.\cite{LayerPRB}  The enhancement relative to the 
analytical approximations persists even beyond separations ($\Delta\approx 
\hbox{4 \AA}$) when there no longer exists an overlap of electron densities.
It arises in part because the complete interaction also contains multipole
interactions.\cite{LayerPRL} 

However, the contrast between the main panel and the insert panel in 
Fig.~\ref{fig:hyperfit} also documents a qualitative change in nature
in the mutual interaction with increasing separation. 
Gradually there is a transition in an intermediate-to-asymptotic regime 
(shown insert panel) where the mutual interaction is essentially specified 
by the morphology of the nanotube density variation as summarized in the
analytical interactions estimate~(\ref{eq:analyticEcnl}).

\section{Conclusions and acknowledgement} We have presented ab 
initio calculations of the binding in nanotube bundles and in nanotube 
dimers. Our calculations rest on a density functional description\cite{vdWGG} 
that includes accounts of the dispersive forces. We have, in addition, presented 
an analytical evaluation valid at intermediate to large nanotube separation.

Our microscopic theory of the CNT binding of semiconducting (8,0) 
CNTs provides a number of results based on the ab initio vdW-DF
calculations. The CNT study supplements recent microscopic theory studies 
of elementents of the DNA base-pair interaction\cite{DNAjacs} and 
of the polyethylene polymer
crystal\cite{PEcrystal} in a broader goal of developing a 
microscopic theory of self-organization and bundling of nanoscale
filaments. 
This vdW-DF study finds a nanotube wall-to-wall separation
in very good agreement with experiments and predicts a vdW bonding with a 
significant strength, consistent with recent measuments for graphitics.\cite{PAHgraph} 

Our work furthermore constitutes an analysis that details the nature of the 
mutual CNT interactions by identifying a set of distinct interaction regimes. 
We provide an analytical approximation for the CNT pair interactions at distances 
when the electron densities are nonoverlapping and the dieletric response are 
dominated by the long-wavelength form. Comparing against our ab initio vdW-DF 
calculations (valid at general distances) we thereby identify a relatively broad 
intermediate-to-asymptotic regime where the interaction form is primarily defined 
by the CNT morphology.

Finally, this introductory study also suggests a framework for an efficient
implementation of quantum-physical modeling of the CNT bundling in more general 
geometries, including nanotube yarn and ropes.  The vdW-DF study documents that a 
summation of nanotube-pair interaction energies represents a fair approximation for 
the nanotube-crystal binding energy when the CNT-pair interaction is calculated in 
vdW-DF. A simple vdW-DF mapping of the nanotube-pair interaction for general 
(independent) CNT rotations relative to the interaction axis therefore 
provides adequate input for describing the vdW bonding in general aligned CNT structures.

We thank B.~I.~Lundqvist for useful discussions and L.~Glasser for suggesting
and detailing the special-function evaluation (\ref{eq:hyperexpress}) from
our corresponding parallel-nanotube interaction result in Ref.~\onlinecite{MSEvdWNT}.
Support from the Swedish Research Council, the Swedish Foundation for 
Strategic Research, the Swedish National Graduate School in Materials Science, 
as well as allocation of computer time at SNIC (Swedish National Infrastructure for Computing) 
is gratefully acknowledged.

\end{document}